\documentclass[11pt, notitlepage, onecolumn, superscriptaddress, nofootinbib, floatfix]{revtex4-1}

\usepackage{amsmath}
\usepackage{mathrsfs}
\usepackage{multirow}
\usepackage{amssymb}
\usepackage{mathtools}
\usepackage{hyperref}
\usepackage[usenames,dvipsnames]{xcolor}
\hypersetup{colorlinks=true, citecolor=Purple, linkcolor=Purple,
urlcolor=Purple}

\usepackage{graphicx}
\graphicspath{{./Figures/}}
\usepackage{orcidlink}

\newcommand{\beq}{\begin{equation}}
\newcommand{\eeq}{\end{equation}}
\newcommand{\ba}{\begin{eqnarray}}
\newcommand{\ea}{\end{eqnarray}}

\definecolor{tclr}{RGB}{148,0,211}

\begin{document}

\allowdisplaybreaks

\title{Parametrized quasinormal modes, greybody factors and their correspondence}

\author{Georgios~Antoniou~\orcidlink{0000-0002-5974-320X}}
\email{georgiosantoniou@tecnico.ulisboa.pt}
\affiliation{CENTRA, Departamento de F\'isica, Instituto Superior T\'ecnico - IST, Universidade de Lisboa - UL, Avenida Rovisco Pais 1, 1049 Lisboa, Portugal}
\affiliation{Dipartimento di Fisica, ``Sapienza'' Universit\'a di Roma, P.A. Moro 5, 00185, Roma, Italy}
\affiliation{Sezione INFN Roma1, P.A. Moro 5, 00185, Roma, Italy}

\begin{abstract}
We present a detailed study of quasinormal modes and greybody factors in the context of the parametrized quasinormal mode framework, in which modifications to general relativity are introduced as small corrections in the potential.
We deduce the QNMs' and GBFs' dependence on the order of the modifications and their polynomial power.
We also test the validity of the recently proposed QNM-GBF correspondence in the pQNM framework by inspecting the regime at which it breaks down.
\end{abstract}

\maketitle


\section{Introduction}
With the advent of gravitational-wave (GW) astronomy, we witness the transformation of black holes (BHs) from theoretical laboratories into precision testbeds of strong-field gravity.
With steadily improving detector sensitivities, next-generation gravitational-wave detectors on the way, and growing event catalogs, we are provided with a unique chance of probing general relativity (GR) at unprecedented accuracy.

Ringdown “spectroscopy” pertains to the study of the last phase of a BH coalescence, in which the remnant BH resembles a ringing bell that settles by emitting a series of damped sinusoids called quasinormal modes (QNMs) \cite{Regge:1957td,Vishveshwara:1970zz,Chandrasekhar:1975zza,Kokkotas:1999bd,Nollert:1999ji,Berti:2009kk}. The latter are complex frequencies which can encode crucial information about the background geometry.
In GR, the QNMs are fully described by the $(n,\ell,m)$ triplet where $n$ denotes the overtone, $\ell$ denotes the multipole number and $m$ is the azimuthal number.

To study QNM deviations from their GR counterparts one may follow a theory-specific prescription which can in principle be quite involved, depending on the modified gravity studied.
This allows for the study of non-perturbative effects but can introduce new degrees of freedom and higher derivative equations of motion which become much harder to handle, let alone decouple.
On the other hand, one may circumvent these difficulties by exploring modified gravity under small perturbations in the beyond-GR contributions.
A great advantage of this approach is that theory-agnostic recipes can be developed which can cover a big parameter space of beyond-GR modifications.
Such an agnostic framework was developed for static and rotating configurations in \cite{Cardoso:2019mqo,McManus:2019ulj,Konoplya:2024lir,Franchini:2022axs,Cano:2024jkd}.
In this so-called parametrized QNM framework (pQNM) the potential is deformed in terms of small, radius-dependent corrections that induce shifts in the QNM spectra.
Originally, a formulation to treat decoupled systems with spherically symmetric perturbations was developed \cite{Cardoso:2019mqo}, while subsequent work extended the formalism to quadratic order and coupled systems \cite{McManus:2019ulj}.
More recent works further adapted pQNM to higher overtones \cite{Hirano:2024fgp}, and performed time-domain computations demonstrating agreement with the frequency-domain analysis \cite{Thomopoulos:2025nuf}.

In parallel, scattering observables, which are called greybody factors (GBFs), quantify the transmission and absorption through the effective potential that governs the wave-propagation around the BH.
In practice, GBFs are derived from solutions of the same equation under different boundary conditions and they correspond to frequency-dependent functions expressing the probability of traversing the potential barrier \cite{Sanchez:1976xm,Sanchez:1977si,Page:1976df}.
Similarly to QNMs, the GBFs can be used as probes for the study of potential GR deviations.
Works studying GBFs in theories beyond GR are fairly limited compared to QNM analyses. For instance, only recently GBFs in scalar-tensor gravity were computed in the non-perturbative regime for a beyond-GR framework \cite{Antoniou:2025bvg}.

Recently, a correspondence between GBFs and QNMs has been derived, where the former are calculated through an approximate relation involving the fundamental mode and the first overtone \cite{Konoplya:2024lir}.
This correspondence has been explored in the case of spherically symmetric potentials and for a special case of rotating BHs \cite{Konoplya:2024vuj}.
In the high-frequency (eikonal) limit the correspondence holds exactly by considering only the fundamental mode, while accuracy is lost when moving to lower multipoles.
Introducing the first overtone, however, was shown to improve the agreement considerably even for lower multipoles.
Closely related analytic developments beyond the strict eikonal
limit were presented in \cite{Konoplya:2023moy}, where compact expressions for both QNMs and GBFs were derived. The QNM-GBF correspondence has since been tested and/or extended for massive fields in Schwarzschild-de Sitter spacetimes \cite{Malik:2024cgb}, quantum-corrected and string-inspired solutions \cite{Skvortsova:2024msa, Dubinsky:2024vbn}, traversable wormholes \cite{Bolokhov:2024otn}, regular black holes with sub-Planckian curvature \cite{Tang:2025mkk}, higher-dimensional black holes \cite{Han:2025cal, Han:2026fpn}, and Kerr geometries \cite{Huang:2025rxx}.

In this work we explore in detail both the QNM spectrum and GBFs of BHs in the pQNM framework.
We present a detailed analysis of the dependence on the radial power of the modifying terms in the effective potential, as well as on the strength of the coupling parameter.
In this context we explore the validity of the correspondence and point out its shortcomings, i.e. the regime of the parameter space where it can no longer be trusted.

We organize this work in the following sections: in Sec. \ref{sec:framework} we introduce the mathematical framework of our analysis, we discuss the pQNM framework and confirm results with existing bibliography, before discussing both QNMs and GBFs in detail.
We push our numerical calculations beyond the point of validity of the pQNM framework to demonstrate the points where it breaks down. 
To ensure the accuracy of our results we employ different numerical approaches and confirm their convergence.
In Sec. \ref{sec:correspondence} we examine the validity of the QNM-GBF correspondence under different numerical approaches.
Finally, we present an outlook in Sec. \ref{sec:outlook}. Supplementary material is provided in the appendices.

\section{Perturbation framework}
\label{sec:framework}

We consider the perturbation of a spin-$s$ field $\Psi_s$ described through the wavelike equation:
\begin{equation}
    \frac{d^2\Psi_s}{dr_*^2}+\left[\omega^2-V_{s}(r_*)\right] \Psi_s=0 \label{eq:Schrodinger}
\end{equation}
where $V_s$ is the perturbation potential of the field $\Psi_s$ and $\omega$ is generally a complex number. In what follows, we will be omitting the spin index unless it is explicitly important\footnote{In the main part of this work we will consider axial $s=2$ perturbations, while in the appendix we present results for $s=0$.}.
For a curved, spherically-symmetric, spacetime background the tortoise coordinate is defined through $dr_* = \sqrt{-\tfrac{g_{rr}}{g_{tt}}} dr$
~\cite{Konoplya:2006rv}.
One may study Eq.~\eqref{eq:Schrodinger} in order to explore the QNM spectrum induced by background or test-field perturbations\footnote{In scalar-tensor gravity for example, the scalar field of the theory is perturbed along the metric, so one can calculate scalar perturbations in addition to the usual gravitational ones. On top of those, one may also explore test field perturbations which yield their distinct QNM spectrum}.
Then the frequency $\omega$ appearing in~\eqref{eq:Schrodinger} is complex, with the real part denoting the oscillation frequency and the imaginary part corresponding to the decay rate.
To search for QNMs, one has to solve \eqref{eq:Schrodinger} with the appropriate boundary conditions, which in this case correspond to purely outgoing waves at infinity and purely ingoing waves at the horizon:
\begin{align}
    \Psi=& \,e^{i \omega r_*}\,, \quad \,\,\,r_* \rightarrow +\infty\,, \nonumber \\
   \Psi=& \,e^{-i \omega r_*}\,, \quad r_* \rightarrow -\infty\,.\label{eq:QNM}
\end{align}
On the other hand, Eq. \eqref{eq:Schrodinger} can be used to study the scattering process of a field due to the gravitational potential in the BH vicinity.
Specifically, the boundary conditions for the aforementioned problem read
\begin{align}
    \Psi=& \,e^{-i \Omega r_*} + R(\Omega)\,e^{i \Omega r_*}\,, \quad r_* \rightarrow +\infty\,, \nonumber \\
   \Psi=& \,A(\Omega)\,e^{-i \Omega r_*}\,, \quad \qquad \quad \,\,r_* \rightarrow -\infty\,,\label{eq:GBF}
\end{align}
where $R(\Omega)$ and $A(\Omega)$ are the reflection and absorption (or transmission) coefficients, respectively and the frequency (now denoted with a capital $\Omega)$ is real.
We have also normalized the parameters in terms of the amplitude of the incoming wave at infinity. We then define the greybody factor (GBF) as
\begin{equation}
    \Gamma_\ell(\Omega) \equiv \,|A_{\ell} (\Omega)|^2 = 1- |R_{\ell}(\Omega)|^2\,.
\end{equation}
Solving Eq.~\eqref{eq:Schrodinger} with the boundary conditions~\eqref{eq:GBF} allows us to determine the greybody factor $\Gamma_{\ell}(\Omega)$ with angular number $\ell$ and frequency $\Omega$.

\subsection{Parametrized framework}

In the pQNM framework our main goal is to match phenomenological modifications of the perturbation potential, with deformations in the QNM spectrum.
The modifications are introduced as $1/r$ corrections in the effective potential and their magnitude is controlled by a bookkeeping parameter.
We will focus on the single-field case of the parametrized framework. The master equation can be written in terms of the perturbed field as
\begin{align}\label{eq:def_PF}
    f\frac{d}{dr}\left( f \frac{d \phi}{dr}  \right) + \left[ \omega^2 - f \tilde{V} \right]{\phi} = 0 \,,
\end{align}
where $f=1-2M/r$, is the metric element for a Schwarzschild background solution.
The potential is decomposed into the GR part (i.e. Regge-Wheeler and Zerilli potentials) and the deviation which is denoted by $\delta V$.
Specifically, we have
\begin{align}
    \tilde{V} =& \tilde{V}_\mathrm{GR} + \delta \tilde{V} \quad , \quad
    \delta \tilde{V} = \frac{1}{r_h^2} \sum_{i=0}^{i_\text{max}} \alpha^{(i)}\left(\frac{r_h}{r}\right)^i \,.
    \label{eq:dV}
\end{align}
The scalar and Regge-Wheeler potentials (axial metric perturbations) are given by:
\begin{align}
\tilde{V}_\mathrm{scalar}  = \frac{\ell(\ell+1)}{r^2} + \frac{2M}{r^3}\quad , \quad
\tilde{V}_\mathrm{RW}  = \frac{\ell(\ell+1)}{r^2} - \frac{6M}{r^3}\,,
\end{align}
and $\delta V(r)$ is the modification to the GR potential\footnote{Notice the unusual convention, which is motivated by existing bibliography on the topic, of not including $f$ in the definition of the potential term above.}.  
The modifications introduced are characterized by the power $i$ of the modification and their corresponding amplitude $\alpha^{(i)}$.
The magnitude of the modification is controlled by the condition \cite{McManus:2019ulj}
\begin{align}\label{eq:criterion}
    \alpha^{(i)} \ll \alpha^{(i)}_\text{max}= (1+1/i)^i(i+1)\,.
\end{align}
\begin{figure}
    \centering
    \includegraphics[width=\linewidth]{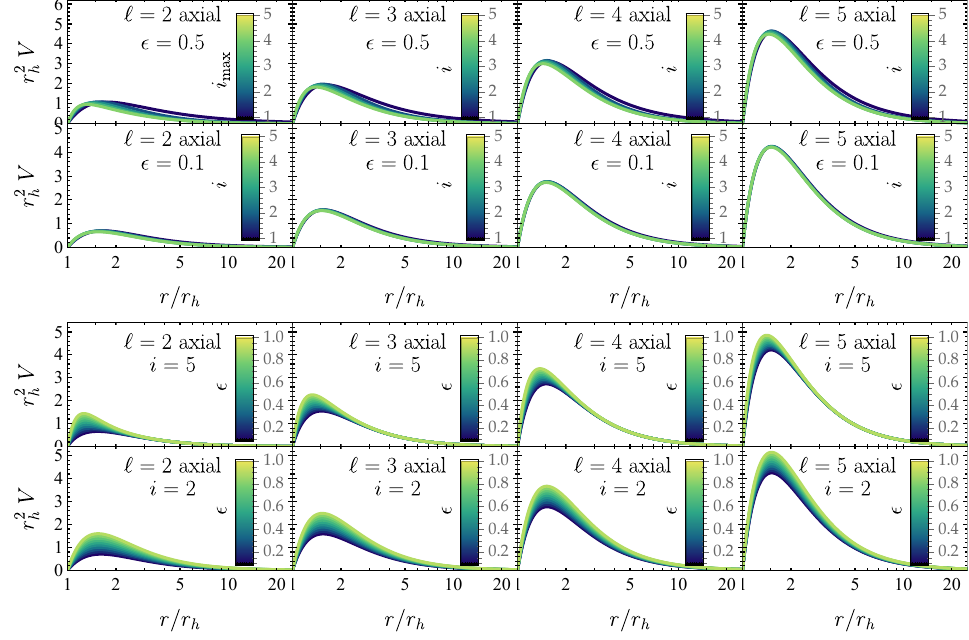}
    \caption{Axial effective potential under deformations.
    \textit{Top:} We fix the bookkeeping parameter at $\epsilon=[0.1,0.5]$ and we vary $i$ from 1 to 5. We plot the results for $\ell=[2,3,4,5]$.
    \textit{Bottom:} Here we fix the power of the modification $i=[2,5]$ and we vary the bookkeeping parameter $\epsilon$ for $\ell=[2,3,4,5]$.}
    \label{fig:potential}
\end{figure}
In Fig. \ref{fig:potential} we show how the potential is modified in the presence of the perturbative corrections \eqref{eq:dV}, by parametrizing in terms of a single bookkeeping parameter $\epsilon$, defined from $\alpha^{(i)}=\epsilon \alpha^{(i)}_\text{max}$.
In the main part of the paper we focus on axial perturbations, but the corresponding analysis for scalar test fields is included in the appendix. Note also that polar perturbations, even though not considered here, should be qualitatively similar.
The purpose of this depiction is to demonstrate the impact of the modifications on the potential and that is why we ignore 0-th order corrections in \eqref{eq:dV} to ensure a vanishing potential at infinity. 
We show the potential deformations for the axial case with multipole numbers $\ell=[2,3,4,5]$, first by fixing the bookkeeping parameter at the values $\epsilon=[0.1,0.5]$ and changing the power $i$ of the modification in the range $i=[1,2,3,4,5]$.
The results are shown in the top panel of Fig. \ref{fig:potential}, where we notice that increasing either the multipole number or $i$ results in a sharper peak for the potential without significantly affecting its maximum height for different $i$.
As expected, the deviations from GR are larger for higher values of $\epsilon$.

We then fix the maximum expansion order at $i=[2,5]$ and we vary the bookkeeping parameter in the range $\epsilon=[0,1]$ with $\delta \epsilon =0.2$. As expected increasing the parameter leads to more pronounced deformations in the potential and higher peaks. Moreover, increasing the power of the modification yields sharper peaks.
In the following sections we will associate the potential deformations appearing here with the corresponding shifts in QNMs and GBFs.

\subsection{Quasinormal modes}
As already explained to find the QNMs we solve \eqref{eq:Schrodinger} with boundary conditions corresponding to ingoing/outgoing waves at the horizon/infinity \eqref{eq:QNM}.
In principle, one expects that small potential deformations should yield small shifts in the QNM spectrum. The modified QNMs' deformation with respect to the GR modes can be expressed as
\begin{equation}
    \omega= \omega_\text{GR}+\sum_{i=0}^{i_\text{max}}\alpha^{(i)}e^{(i)}\, .
    \label{eq:omega_sum}
\end{equation}
where $M\omega_\text{GR}$ denotes the usual GR mode (i.e. for $(n,\ell)=(0,2)$ we have $M\omega_\text{GR}\approx 0.3637-0.08896 i$).
It is important to note that the coefficients $e^{(i)}$, even though dependent on the type of the field under perturbation, do not depend on the amplitude of the modification $\alpha^{(i)}$.
The values of the coefficients have already been computed in various works \cite{Cardoso:2019mqo,McManus:2019ulj,Hirano:2024fgp,Volkel:2022aca,Cano:2024jkd}.
Using Leaver's method \cite{Leaver:1985ax} we confirm results with bibliography regarding the coefficients in the QNM frequency expansions, by calculating the first few ones for the fundamental axial modes with $\ell=2$. Our results have been obtained with a $N=200$ step continued fraction approach where we applied Gaussian elimination steps in order to yield a 3-term recurrence relation of the form
\begin{align}
&{\alpha}_0\, a_1+{\beta}_0 \, a_0=0\,,\\
&{\alpha}_n \, a_{n+1}+{\beta}_n \, a_n+{\gamma}_n\, a_{n-1}=0\,,\qquad n>0\,,
\end{align}
where the coefficients $\alpha_n,\,\beta_n,\,\gamma_n$ appearing above depend on the background (their full expressions are given in the Appendix), and the expressions $a_n$ are used to construct the analytical expression of the perturbation function, namely
\begin{equation}
    \phi(r)\sim \left(1-\frac{2M}{r}\right)^{-4iM\omega}\left(\frac{r}{2M}-1\right)^{2iM\omega}e^{i\omega r}\sum_n a_n^{(i)} \left(1-\frac{2M}{r}\right)^{n}\,.
\end{equation}
This provides an additional independent confirmation of the established results. However, since we only compute a few representative coefficients (only for the fundamental mode) in what follows we make use of the values provided in \cite{Hirano:2024fgp} for $n=0$ and $n=1$.

\begin{figure}
    \centering
    \includegraphics[width=\linewidth]{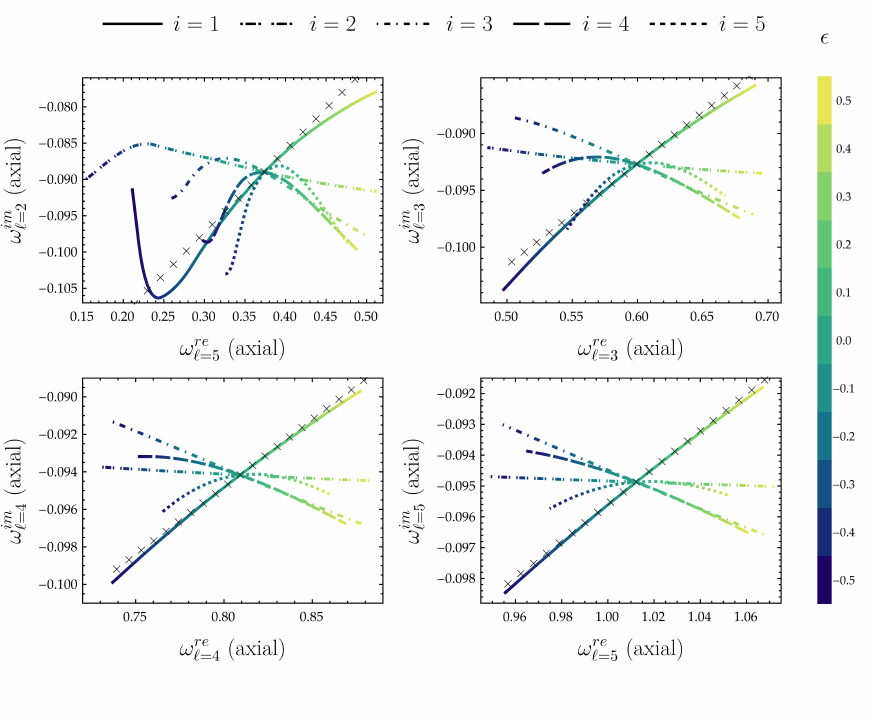}
    \caption{
    We show the real and imaginary parts of the axial fundamental ($n=0$) QNMs with angular numbers $\ell=[2,3,4,5]$ and a modified potential of power $i=[1,2,3,4,5]$, while varying the parameter $\epsilon$ in the range $[-0.5,0.5]$.
    The colour gradient tracks changes with respect to $\epsilon$ and each dashing corresponds to a different power $i$ for the potential modification.
    }
    \label{fig:QNMs_axial}
\end{figure}

Since one of our goals is to map out the regime of validity of the pQNM framework, we also proceed to compute the QNMs using a direct-integration (DI) method.
Specifically, we solve \eqref{eq:Schrodinger} by employing a series expansion at the horizon and infinity
\begin{equation}
\lim_{r\to 2M}\phi\sim\;(r-2M)^{-2iM\omega}\sum_{n=0} \phi_n (r-2M)^{n}\quad , \quad
\lim_{r\to \infty}\phi \sim \; r^{2iM\omega} e^{i\omega r} \sum_{n=0} \frac{\phi_n}{r^n}\, .
\end{equation}
For the DI integration results we found it sufficient to consider expansions up to order $n=5$. However, for all results produced via this approach in this work, we considered $n=7$ to secure the numerical stability of our results.
In Fig. \ref{fig:QNMs_axial} we show the real and imaginary parts of the axial QNMs. Each one of the four plots corresponds to a different multipole number, namely $\ell=[2,3,4,5]$. Each line pertains to a different value $i$ for the potential modification in the range $i=[1,2,3,4,5]$ and every color of the color gradient is associated with a different value for the parameter $\epsilon$. We keep the parameter $\epsilon$ at small values, namely $\epsilon\in [-0.5,0.5]$ to remain within -or at least close to- the regime of validity of the pQNM framework.

To help track the changes in the parameter $\epsilon$ we employ the aforementioned colour gradient where the yellow limit corresponds to the maximum value of $\epsilon$ ($\epsilon_{\text{max}}=+0.5$) and the blue to the lowest one ($\epsilon_{\text{min}}=-0.5$).
Overall, we deduce that taking algebraically smaller values for $\epsilon$ reduces the oscillation frequency for all multipole numbers.
The real part of the QNM appears to be similar for fixed $\epsilon$ for all curves.
This is consistent with the behaviour of the effective potential in the top panel of Fig. \ref{fig:potential}, where the potential height appeared relatively unchanged.
The behaviour of the imaginary part is generally more involved and can even become non-monotonic.
Also in agreement with the lower panel of Fig. \ref{fig:potential}, the oscillation frequency increases while increasing the bookkeeping parameter for a certain modified potential.
In terms of the multipole number $\ell$ we notice that while increasing it, the changes of the QNMs with respect to the GR limit become smaller, while the non-monotonic behaviour smoothens out for a fixed $\epsilon$.

Moreover, for each one of the four plots in Fig. \ref{fig:QNMs_axial} we show with distinct points, the QNMs we derived from the pQNM formulation for the curves with $i=1$ (we avoid including them for the remaining curves to maintain visual clarity).
As anticipated, increasing $\epsilon$ results in inconsistencies with the non-perturbative results, with the discrepancies subsiding when we increase $\ell$ for a fixed $\epsilon$.
It is also quite clear that the pQNM expansion is unable to capture the non-monotonicities in the curves. This, however, should not be a problem, since by that point we have already pushed our calculations beyond the regime of validity of the pQNM framework since we have moved significantly from the GR limit.

\subsection{Greybody factors}

\begin{figure}
    \centering
    \includegraphics[width=\linewidth]{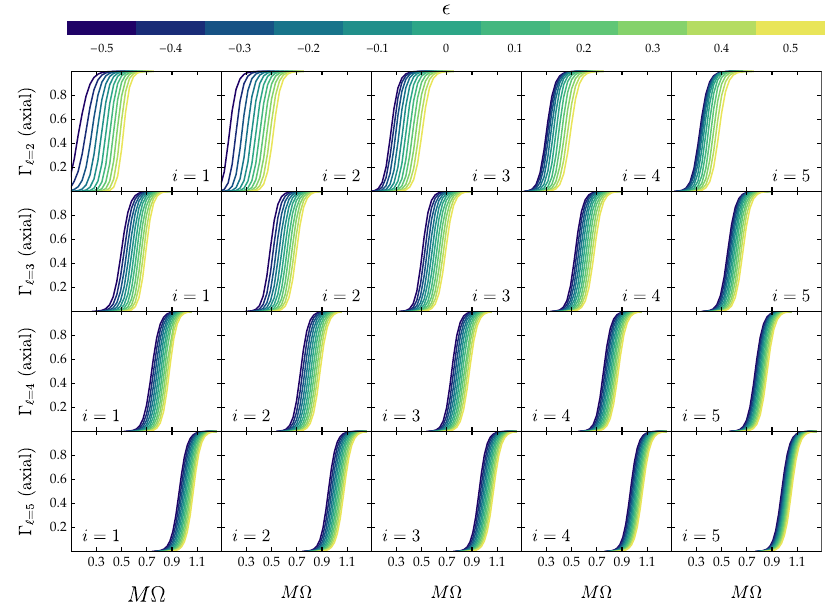}
    \caption{
    Greybody factors for angular numbers $\ell=[2,3,4,5]$ and a modified potential of power $i=[1,2,3,4,5]$, while varying the parameter $\epsilon$ in the range $[-0.5,0.5]$.
    Once again colour gradient tracks changes with respect to $\epsilon$ and is consistent with the one of Fig. \ref{fig:QNMs_axial}.
    }
    \label{fig:GBFs}
\end{figure}

We then proceed to calculate the GBFs for the cases studied in the previous subsection.
As already explained we solve the same equation \eqref{eq:Schrodinger} with different boundary conditions \eqref{eq:GBF} while we employ a consistent precision with the integrators we used to calculate the QNMs earlier by following a DI approach.
We present the results in Fig. \ref{fig:GBFs} where each row is characterized by a different multipole number $\ell$ and each column corresponds to a different power $i$ in the modified potential.
Again the colour gradient tracks changes in $\epsilon$ and is consistent with the one we used for Fig. \ref{fig:QNMs_axial}, thus allowing a direct comparison.

The first thing we notice for all different choices for $(\ell,i,\epsilon)$ is that the real part of the QNM frequencies appearing in Fig. \ref{fig:QNMs_axial} corresponds to the regime of $\Omega$ where the GBF transitions from zero to one, an effect consistent with what is observed in usual GR analyses.
The above is reflected on the fact that for all multipole numbers increasing the power of the modified potential leads to a wider spread for the transition $i_\text{max}=1\to i_\text{max}=2$ and to a narrower one for $i_\text{max}=(2,3,4)\to i_\text{max}=(3,4,5)$ in agreement with the span of the real QNM part of each curve depicted in Fig. \ref{fig:QNMs_axial}.
Moreover, in all cases increasing the value of the bookkeeping parameter leads to the GBF curves shifting to larger frequencies, which is consistent with the monotonicity of the real part of the QNMs shown in Fig. \ref{fig:QNMs_axial}.


\section{QNM-GBF correspondence}
\label{sec:correspondence}

Recently, as explained in the introduction of this work, a correspondence between QNMs and GBFs has been established for static \cite{Konoplya:2024lir} (see also \cite{Pedrotti:2025idg}) and rotating black holes \cite{Konoplya:2024vuj}.
We will refrain from presenting the exact steps for the derivation of the correspondence but below we will present the basic results.
The correspondence relies on the WKB approach \cite{Schutz:1985km,Iyer:1986np,Iyer:1986nq,Konoplya:2011qq,Konoplya:2019hlu}, according to which one has to match the asymptotic solutions of \eqref{eq:Schrodinger} in an intermediate region in the vicinity of the potential peak.
In that regime the potential can be expanded as $V=\sum_n \frac{d^n V}{dr_*^n}\frac{(r_*-r_0)}{n!}\vert_{r_0}$, where $r_0$ is the tortoise coordinate at the peak.
Consequently \eqref{eq:Schrodinger} takes the form
\begin{equation}
    \frac{d^2}{dr_*^2} \Psi + \left[ \omega^2 - V(r_0) - \frac{1}{2} \frac{d^2 V(r_0)}{dr_*^2} (r_* - r_0)^2 +\ldots\right] \Psi = 0.
\end{equation}
We will refrain from presenting the WKB method in detail here since it has been extensively used in the bibliography, see e.g. \cite{Konoplya:2006rv}.
The WKB formula to arbitrary order is given by:
\begin{equation}
\omega^{2}
= V(r_{0})
+ A_{2}(\mathcal{K}^{2})
+ A_{4}(\mathcal{K}^{2})
+ \ldots
- i\mathcal{K}
\sqrt{-2\frac{d^{2}V(r_{0})}{dr_{*}^{2}}}
\left[ 1 + A_{3}(\mathcal{K}^{2}) + A_{5}(\mathcal{K}^{2}) + \ldots \right]\, ,
\label{eq:WKB}
\end{equation}
where the functions $\mathcal{A}_i$ represent the i-th order WKB correction, and $\mathcal{K}$ is dependent on the boundary conditions considered.
For QNMs $\mathcal{K}=n+1/2$ where $n$ is the overtone number, and the corresponding QNM frequency is denoted with $\omega_n$.
For spherically symmetric backgrounds one may express the effective potential in terms of the multipole number $\ell$ (which has to be greater than or equal to the spin of the perturbation), i.e.:
\begin{equation}
V_{\ell}(r_*) = \ell^{2}U_{0}(r_*)+ \ell\,U_{1}(r_*)+ U_{2}(r_*)+ \ell^{-1}U_{3}(r_*)+ \ldots
\label{eq:Vell}
\end{equation}
This allows us to express the frequency $\omega$ as an expansion in powers of the multipole number, i.e.
\begin{equation}
\omega = \ell \sqrt{U_{0}(r_{0})}- i\mathcal{K}\sqrt{-\frac{d^{2}U_{0}(r_{0})/dr_{*}^{2}}{2U_{0}(r_{0})}}+ \mathcal{O}(\ell^{-1}).
\label{eq:omega_ell}
\end{equation}
On the other hand, the reflection and transmission coefficients of the scattering problem are given in terms of $\mathcal{K}$ as
\begin{align}
    \vert R\vert^2 = & \left(1+e^{-2i\pi\mathcal{K}}\right)^{-1}\, ,\label{eq:Tcoeff}\\
    \vert T\vert^2 = & \left(1+e^{+2i\pi\mathcal{K}}\right)^{-1}\, ,
\end{align}
where $\vert T\vert^2=1-\vert R\vert^2$.
The correspondence formula between QNMs and GBFs is derived by first retrieving the QNMs from \eqref{eq:WKB} in terms of the effective potential and its derivatives and expressing them as an expansion in powers of $\ell$ through \eqref{eq:Vell}.
We then solve \eqref{eq:WKB} for $\mathcal{K}$, substitute the result in \eqref{eq:Tcoeff}, and use the appropriate WKB order to truncate for the desired power of $\ell$.
The result has been derived in \cite{Konoplya:2024lir} to order $\mathcal{O}(\ell^{-2})$, where additional details are provided. Here we present the result where the 6th order WKB has been considered\footnote{To 6th order the WKB approach yields
\begin{equation*}
\mathcal{K} = i\,\frac{\omega^{2} - V_{0}}{\sqrt{-2V_{2}}}
- \sum_{k=2}^{6} \Lambda_{k}(\mathcal{K}),
\end{equation*}
where the higher-order WKB corrections $\Lambda_k$ can be found, see e.g. \cite{Iyer:1986np,Iyer:1986nq,Konoplya:2019hlu}.}
\cite{Konoplya:2024lir}:
\begin{equation}
\begin{aligned}
i\mathcal{K}
=& \;\frac{\Omega^{2}-\left(\omega_0^{\mathrm{re}}\right)^{2}}
{4\,\omega_0^{\mathrm{re}}\,\omega_0^{\mathrm{im}}}
\Bigg[1+ \frac{\left(\omega_0^{\mathrm{re}}-\omega_1^{\mathrm{re}}\right)^{2}}
{32\,\left(\omega_0^{\mathrm{im}}\right)^{2}}
- \frac{3\,\omega_0^{\mathrm{im}}-\omega_1^{\mathrm{im}}}
{24\,\omega_0^{\mathrm{im}}}- \frac{\omega_0^{\mathrm{re}}-\omega_1^{\mathrm{re}}}
{16\,\omega_0^{\mathrm{im}}}
\\
&\,
- \frac{\left(\omega^{2}-\left(\omega_0^{\mathrm{re}}\right)^{2}\right)^{2}}
{16\,\left(\omega_0^{\mathrm{re}}\right)^{3}\omega_0^{\mathrm{im}}}
\Bigg(
1 +
\frac{\omega_0^{\mathrm{re}}\left(\omega_0^{\mathrm{re}}-\omega_1^{\mathrm{re}}\right)}
{4\,\left(\omega_0^{\mathrm{im}}\right)^{2}}
\Bigg)
+ \frac{\left(\omega^{2}-\left(\omega_0^{\mathrm{re}}\right)^{2}\right)^{3}}
{32\,\left(\omega_0^{\mathrm{re}}\right)^{5}\omega_0^{\mathrm{im}}}
\Bigg(
1 +
\frac{\omega_0^{\mathrm{re}}\left(\omega_0^{\mathrm{re}}-\omega_1^{\mathrm{re}}\right)}
{4\,\left(\omega_0^{\mathrm{im}}\right)^{2}}
\Bigg)
\\
&\,
+ \left(\omega_0^{\mathrm{re}}\right)^{2}
\Bigg(
\frac{\left(\omega_0^{\mathrm{re}}-\omega_1^{\mathrm{re}}\right)^{2}}
{16\,\left(\omega_0^{\mathrm{im}}\right)^{4}}
- \frac{3\,\omega_0^{\mathrm{im}}-\omega_1^{\mathrm{im}}}
{12\,\omega_0^{\mathrm{im}}}
\Bigg)
\Bigg]
+ \mathcal{O}\!\left(\frac{1}{\ell^{3}}\right).
\end{aligned}
\label{eq:correspondence}
\end{equation}





\begin{figure}
    \centering
    \includegraphics[width=\linewidth]{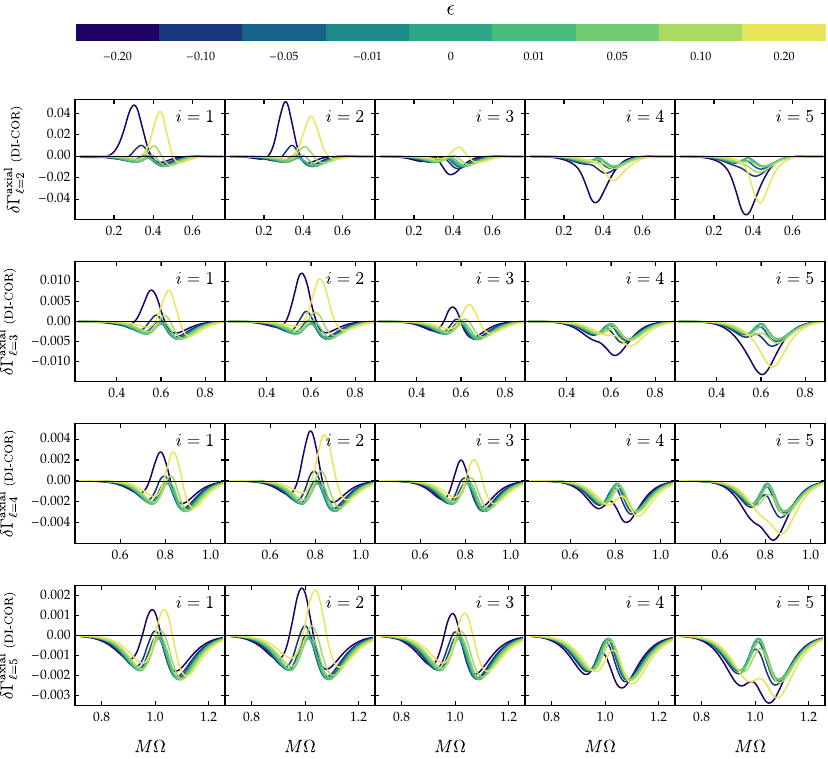}
    \caption{Relative differences between the GBFs derived using a DI approach and the QNM-GBF correspondence.
    We clearly notice better agreement between the two methods while increasing the multipole number $\ell$.
    Overall, in all cases the maximum differences tend to appear close to the point where the GBF transitions from zero to one.}
    \label{fig:DI_COR}
\end{figure}

\begin{figure}
    \centering
    \includegraphics[width=\linewidth]{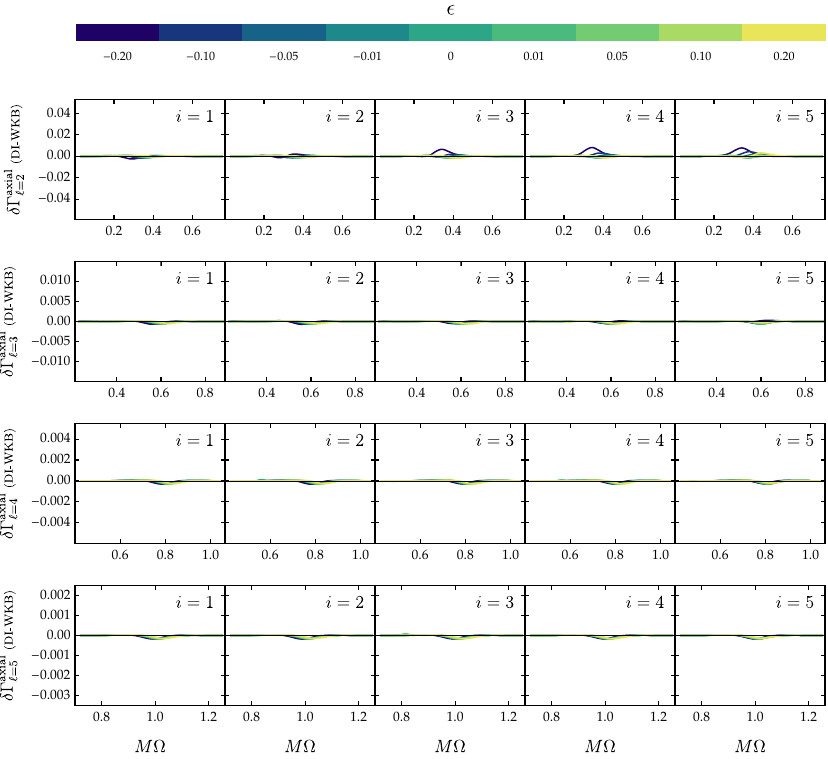}
    \caption{Same as Fig. \ref{fig:DI_COR} but comparing the results of the DI with the 3rd order WKB method.
    We maintain the same scale for each $\ell$ in order to visualize the quantitative differences with Fig. \ref{fig:DI_COR}.
    It is quite clear that even at 3rd order the WKB yields results much closer to those of the numerical integration.
    }
    \label{fig:DI_WKB}
\end{figure}

To quantify the success of the approximate relationship \eqref{eq:correspondence} in capturing the correct GBF behaviour, we plot the difference between the results derived directly from it, as well as from the direct integration approach described in the previous section.
The results are shown in Fig. \ref{fig:DI_COR}.
Each colour of the colour gradient corresponds to different strengths for the coupling $\epsilon$, ranging from $-0.2$ and $+0.2$.
We keep $\epsilon$ at relatively small values in order to remain close to the regime of validity of the pQNM framework.
As expected, increasing the multipole number $\ell$ yields better agreement since the WKB approximation from which \eqref{eq:correspondence} derives, holds more accurately.
Another interesting observation relates to the fact that the deviations are more sensitive to the coupling strength for lower multipoles, while increasing $\vert \epsilon \vert$ always results in larger deviations.
Moreover, the deviations appear to be slightly larger for negative $\epsilon$.

In order to fully assess the validity of the correspondence and the extent to which it holds or breaks down due to the WKB approximation or the multipole number truncation, we also calculated the GBF differences $\delta\Gamma$ between DI and the 3rd order WKB.
We show the results in Fig. \ref{fig:DI_WKB} maintaining the same scale both in the vertical and horizontal axes with respect to Fig. \ref{fig:DI_COR}, to allow for a direct comparison.
It is quite evident that for any multipole $\ell$ or power $i$ the quantities $\delta\Gamma_{\text{DI-COR}}$ are orders of magnitude larger than $\delta\Gamma_{\text{DI-WKB}}$ even though the latter are calculated with the 3rd order WKB formulas.

At this point it is worth understanding the type of errors that enter the calculations when one uses WKB versus the correspondence approximations.
The order of the WKB is reflected on the terms we consider on the RHS of \eqref{eq:WKB}.
At 3rd WKB order, the quantity $\mathcal{K}$ is determined in terms of derivatives of the potential up to $V^{(6)}$, while at 6th order we consider derivatives up to $V^{(12)}$.
These derivatives contain important information about the shape of the potential, part of which may be lost when one takes the correspondence approximation which makes use of the first two modes calculated after truncating in terms of the multipole number.
In this sense making use of $\eqref{eq:correspondence}$ results in taking an ``approximation of an approximation'', which is incapable of always capturing the intricacies of the shape of the potential, in the same way that the straightforward WKB method allows.

\section{Discussion}
\label{sec:outlook}

In this work we performed a comprehensive analysis of QNMs and GBFs in the context of the pQNM framework, where GR modifications are introduced as additional inverse polynomial terms in the effective potential.
By specifically studying the QNM spectrum we deduced the regime of validity of the pQNM framework by extending the coupling strengths beyond the perturbative limit, concluding that it is safe to use it with significant precision when $\epsilon < 0.1$.
We then calculated the GBFs for various multipole numbers $\ell$ and powers for the potential-modifying terms, showing that higher multipoles result in smaller relative deviations from the GR results.

In testing the QNM-GBF correspondence which relies on using the 6th order WKB to truncate up to $\mathcal{O}(\ell^{-2})$, we found satisfactory consistency when one calculates GBF for higher multipoles.
This is not surprising as the correspondence relies on the substitution of the first two modes, which do not always manage to capture the shape of the potential and its derivatives in a sufficient way.
It is worth pointing out that extending the correspondence to higher multipoles will not necessarily improve the agreement, since the higher overtones that will be involved are generally less reliably captured with barrier-top WKB methods and are more sensitive to truncation errors in the perturbative pQNM expansions.
The exploration of such a scenario is left for future work.
We saw that, in practice, relying on a lower WKB approximation (3rd order in our case) manages to capture the GBF behavior better than the correspondence (which relies on 6th order WKB).
This is related to the fact that WKB is applied on the full potential, whereas the correspondence relies on reconstructing the GBFs from the first two QNMs with truncation with respect to $\ell^{-1}$, and thus loses intricate information about the shape of the potential.

Our analysis suggests that there exists domain along which combining the results of the pQNM framework for the first two modes and subsequently using the correspondence to produce the GBF curves, yields results that are very close to the ones derived with the full numerical integration. However, this region of the parameter space requires one to consider high multipoles and remain perturbatively close to the GR results.

\acknowledgements
The author acknowledges financial support provided by FCT - Fundação para a Ciência e a Tecnologia, I.P., through the ERC-Portugal program Project ``GravNewFields'' and also thanks the Fundação para a Ciência e Tecnologia (FCT), Portugal, for the financial support to the Center for Astrophysics and Gravitation (CENTRA/IST/ULisboa) through grant No.~\href{https://doi.org/10.54499/UID/PRR/00099/2025}{UID/PRR/00099/2025} and grant No.~\href{https://doi.org/10.54499/UID/00099/2025}{UID/00099/2025}.
This work was also supported by the INFN TEONGRAV initiative.

\appendix

\section{Recurrence relations}
For the continued fraction approach we employ in order to derive the pQNM framework coefficients, we had to first reduce a 4-term recurrence relation to a 3-term one by applying Gaussian elimination.
Specifically, for $i=1,2,3$ we have:
\begin{align}
&{\alpha}_0 {U}_1+{\beta}_0 {U}_0=0\,,\\
&{\alpha}_1 {U}_{2}+{\beta}_1 {U}_1+{\gamma}_1 {U}_{0}=0\,,\\
&{\alpha}_2 {U}_{3}+{\beta}_2 {U}_2+{\gamma}_2 {U}_{1}+{\delta}_2 {U}_{0}=0\,,\\
&{\alpha}_n {U}_{n+1}+{\beta}_n {U}_n+{\gamma}_n {U}_{n-1}+{\delta}_n {U}_{n-2}+{\epsilon}_n {U}_{n-3}=0\,,\qquad n>2\,,
\end{align}
which can be recast into
\begin{align}
&{\alpha}_0^{(2)} {U}_1+{\beta}_0^{(2)} {U}_0=0\,,\\
&{\alpha}_n^{(2)} {U}_{n+1}+{\beta}_n^{(2)} {U}_n+{\gamma}_n^{(2)} {U}_{n-1}=0\,,\qquad n>0\,,
\end{align}
where the second order coefficients are given in terms of the first order ones by
\begin{align}
&{\alpha}_n^{(2)}={\alpha}_n^{(1)},\\
&{\beta}_n^{(2)}={\beta}_n^{(1)}-{\delta}_n^{(1)}\left[ {\gamma}_{n-1}^{(2)} \right]^{-1} {\alpha}_{n-1}^{(2)},\\
&{\gamma}_n^{(2)}={\gamma}_n^{(1)}-{\delta}_n^{(1)}\left[{\gamma}_{n-1}^{(2)}\right]^{-1} {\beta}_{n-1}^{(2)},
\end{align}
and the first order ones in terms of the zeroth order coefficients by
\begin{align}
&{\alpha}_n^{(1)}={\alpha}_n^{(0)},\\
&{\beta}_n^{(1)}={\beta}_n^{(0)}-{\epsilon}_n^{(0)}\left[ {\delta}_{n-1}^{(1)} \right]^{-1} {\alpha}_{n-1}^{(1)},\\
&{\gamma}_n^{(1)}={\gamma}_n^{(0)}-{\epsilon}_n^{(0)}\left[{\delta}_{n-1}^{(1)}\right]^{-1} {\beta}_{n-1}^{(1)},\\
&{\delta}_n^{(1)}={\delta}_n^{(0)}-{\epsilon}_n^{(0)}\left[{\delta}_{n-1}^{(1)} \right]^{-1}{\gamma}_{n-1}^{(1)}\, .
\end{align}
For higher power corrections the terms in the recurrence relations increase, e.g. for $i=4,5$ we have a $6$ or $7$-term recurrence relation requiring further elimination steps.
The QNM frequencies are solutions to the equation ${M}{U}_0=0$, where ${M}\equiv {\beta}_0+{\alpha}_0 {R}^{\dagger}_0$ with ${U}_{n+1}={R}^{\dagger}_n {U}_n$ and ${R}^{\dagger}_n=-\left({\beta}_{n+1}+{\alpha}_{n+1}{R}^{\dagger}_{n+1}\right)^{-1} {\gamma}_{n+1}$.

\section{Scalar case}

\begin{figure}
    \centering
    \includegraphics[width=\linewidth]{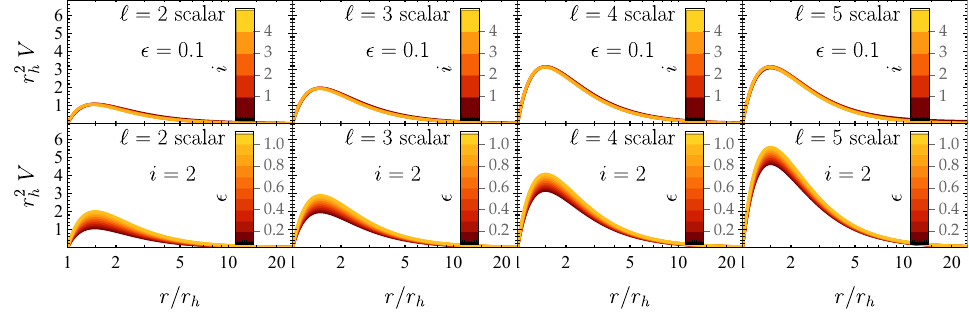}\\
    \caption{Scalar effective potential under deformations.
    In the first row, we fix the bookkeeping parameter at $\epsilon=0.1$ and we vary $i$ from 1 to 5. We plot the results for $\ell=[2,3,4,5]$.
    In the second row, we fix the power of the modification $i=2$ and we vary the bookkeeping parameter $\epsilon$, with $\ell=[2,3,4,5]$.}
    \label{fig:scalar_V}
\end{figure}

In this appendix, and in order to complement the analysis of the axial perturbations in the main part of the paper, we present results pertaining to the scalar sector.
Overall, the differences are mostly qualitative and the results we draw regarding the QNMs and GBFs as well as the correspondence between the two are similar.
In Fig. \ref{fig:scalar_V} we demonstrate the scalar potential's dependence on $i$ and the coupling strength for different $\ell$. We notice a sharper peak for higher multipoles and deviations becoming larger when one increases $\epsilon$.

\begin{figure}
    \centering
    \includegraphics[width=\linewidth]{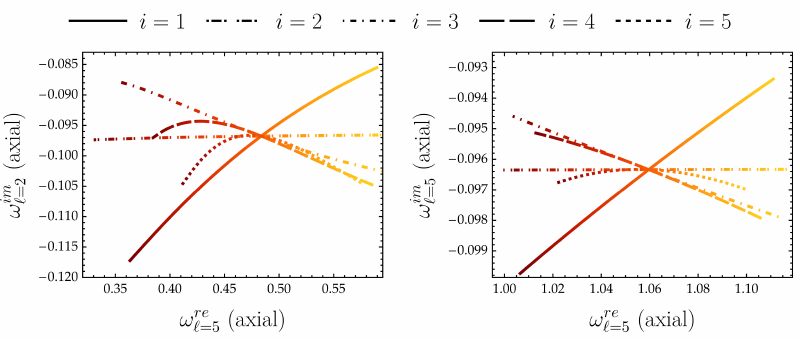}\\
    \includegraphics[width=\linewidth]{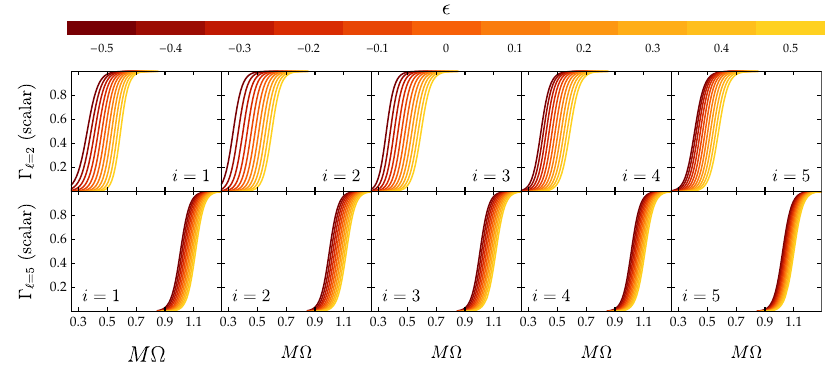}
    \caption{
    \textit{Top:} we show the real and imaginary parts of the scalar fundamental ($n=0$) QNMs with angular numbers $\ell=[2,5]$ and a modified potential of power $i=[1,2,3,4,5]$, while varying the parameter $\epsilon$ in the range $[-0.5,0.5]$.
    The colour gradient tracks changes with respect to $\epsilon$ and each dashing corresponds to a different power $i$ for the potential modification.
    \textit{Bottom:} Greybody factors for angular numbers $\ell=[2,5]$ and a scalar modified potential of power $i=[1,2,3,4,5]$, while varying the parameter $\epsilon$ in the range $[-0.5,0.5]$.
    The colour gradient is consistent with the one from the top panel.
    }
    \label{fig:QNMs_GBFs_scalar}
\end{figure}

In Fig. \ref{fig:QNMs_GBFs_scalar} we present the QNMs and GBFs for scalar perturbations, for $i=[1,2,3,4,5]$ and $\ell=[2,5]$.
We use a matching color scheme to allow for direct visual comparisons.
The results are qualitatively similar to those obtained for the axial perturbations.

\bibliography{bibnote}

@article{Thomopoulos:2025nuf,
    author = {Thomopoulos, Spyros and V{\"o}lkel, Sebastian H. and Pfeiffer, Harald P.},
    title = "{Ringdown spectroscopy of phenomenologically modified black holes}",
    eprint = "2504.17848",
    archivePrefix = "arXiv",
    primaryClass = "gr-qc",
    doi = "10.1103/xtzl-lyn6",
    journal = "Phys. Rev. D",
    volume = "112",
    number = "6",
    pages = "064054",
    year = "2025"
}

@article{Hirano:2024fgp,
    author = "Hirano, Shin'ichi and Kimura, Masashi and Yamaguchi, Masahide and Zhang, Jiale",
    title = "{Parametrized black hole quasinormal ringdown formalism for higher overtones}",
    eprint = "2404.09672",
    archivePrefix = "arXiv",
    primaryClass = "gr-qc",
    reportNumber = "RUP-24-6",
    doi = "10.1103/PhysRevD.110.024015",
    journal = "Phys. Rev. D",
    volume = "110",
    number = "2",
    pages = "024015",
    year = "2024"
}

@article{Cardoso:2019mqo,
    author = "Cardoso, Vitor and Kimura, Masashi and Maselli, Andrea and Berti, Emanuele and Macedo, Caio F. B. and McManus, Ryan",
    title = "{Parametrized black hole quasinormal ringdown: Decoupled equations for nonrotating black holes}",
    eprint = "1901.01265",
    archivePrefix = "arXiv",
    primaryClass = "gr-qc",
    doi = "10.1103/PhysRevD.99.104077",
    journal = "Phys. Rev. D",
    volume = "99",
    number = "10",
    pages = "104077",
    year = "2019"
}

@article{McManus:2019ulj,
    author = "McManus, Ryan and Berti, Emanuele and Macedo, Caio F. B. and Kimura, Masashi and Maselli, Andrea and Cardoso, Vitor",
    title = "{Parametrized black hole quasinormal ringdown. II. Coupled equations and quadratic corrections for nonrotating black holes}",
    eprint = "1906.05155",
    archivePrefix = "arXiv",
    primaryClass = "gr-qc",
    doi = "10.1103/PhysRevD.100.044061",
    journal = "Phys. Rev. D",
    volume = "100",
    number = "4",
    pages = "044061",
    year = "2019"
}

@article{Konoplya:2024lir,
    author = "Konoplya, R. A. and Zhidenko, A.",
    title = "{Correspondence between grey-body factors and quasinormal modes}",
    eprint = "2406.11694",
    archivePrefix = "arXiv",
    primaryClass = "gr-qc",
    doi = "10.1088/1475-7516/2024/09/068",
    journal = "JCAP",
    volume = "09",
    pages = "068",
    year = "2024"
}

@article{Franchini:2022axs,
    author = {Franchini, Nicola and V{\"o}lkel, Sebastian H.},
    title = "{Parametrized quasinormal mode framework for non-Schwarzschild metrics}",
    eprint = "2210.14020",
    archivePrefix = "arXiv",
    primaryClass = "gr-qc",
    doi = "10.1103/PhysRevD.107.124063",
    journal = "Phys. Rev. D",
    volume = "107",
    number = "12",
    pages = "124063",
    year = "2023"
}

@article{Cano:2024jkd,
    author = {Cano, Pablo A. and Capuano, Lodovico and Franchini, Nicola and Maenaut, Simon and V{\"o}lkel, Sebastian H.},
    title = "{Parametrized quasinormal mode framework for modified Teukolsky equations}",
    eprint = "2407.15947",
    archivePrefix = "arXiv",
    primaryClass = "gr-qc",
    doi = "10.1103/PhysRevD.110.104007",
    journal = "Phys. Rev. D",
    volume = "110",
    number = "10",
    pages = "104007",
    year = "2024"
}

@article{Konoplya:2006rv,
    author = "Konoplya, R. A. and Zhidenko, A.",
    title = "{Perturbations and quasi-normal modes of black holes in Einstein-Aether theory}",
    eprint = "gr-qc/0605082",
    archivePrefix = "arXiv",
    doi = "10.1016/j.physletb.2006.11.036",
    journal = "Phys. Lett. B",
    volume = "644",
    pages = "186--191",
    year = "2007"
}

@article{Tang:2025mkk,
    author = "Tang, Chen and Ling, Yi and Jiang, Qing-Quan",
    title = "{Correspondence between grey-body factors and quasinormal modes for regular black holes with sub-Planckian curvature*}",
    eprint = "2503.21597",
    archivePrefix = "arXiv",
    primaryClass = "gr-qc",
    doi = "10.1088/1674-1137/adfa74",
    journal = "Chin. Phys.",
    volume = "49",
    number = "12",
    pages = "125110",
    year = "2025"
}

@article{Konoplya:2024vuj,
    author = "Konoplya, R. A. and Zhidenko, A.",
    title = "{Correspondence between grey-body factors and quasinormal frequencies for rotating black holes}",
    eprint = "2408.11162",
    archivePrefix = "arXiv",
    primaryClass = "gr-qc",
    doi = "10.1016/j.physletb.2025.139288",
    journal = "Phys. Lett. B",
    volume = "861",
    pages = "139288",
    year = "2025"
}

@article{Pedrotti:2025idg,
    author = "Pedrotti, Davide and Calz{\`a}, Marco",
    title = "{Trinity of black hole correspondences: Shadows, quasinormal modes, graybody factors, and cautionary remarks}",
    eprint = "2504.01909",
    archivePrefix = "arXiv",
    primaryClass = "gr-qc",
    doi = "10.1103/1q35-mjjz",
    journal = "Phys. Rev. D",
    volume = "111",
    number = "12",
    pages = "124056",
    year = "2025"
}

@article{Konoplya:2019hlu,
    author = "Konoplya, R. A. and Zhidenko, A. and Zinhailo, A. F.",
    title = "{Higher order WKB formula for quasinormal modes and grey-body factors: recipes for quick and accurate calculations}",
    eprint = "1904.10333",
    archivePrefix = "arXiv",
    primaryClass = "gr-qc",
    doi = "10.1088/1361-6382/ab2e25",
    journal = "Class. Quant. Grav.",
    volume = "36",
    pages = "155002",
    year = "2019"
}

@article{Konoplya:2011qq,
    author = "Konoplya, R. A. and Zhidenko, A.",
    title = "{Quasinormal modes of black holes: From astrophysics to string theory}",
    eprint = "1102.4014",
    archivePrefix = "arXiv",
    primaryClass = "gr-qc",
    doi = "10.1103/RevModPhys.83.793",
    journal = "Rev. Mod. Phys.",
    volume = "83",
    pages = "793--836",
    year = "2011"
}

@article{Antoniou:2025bvg,
    author = "Antoniou, Georgios and Pappas, Thomas D. and Kanti, Panagiota",
    title = "{Greybody factors in scalar-tensor gravity and beyond}",
    eprint = "2507.17329",
    archivePrefix = "arXiv",
    primaryClass = "gr-qc",
    doi = "10.1103/zwhl-sqqs",
    journal = "Phys. Rev. D",
    volume = "112",
    number = "8",
    pages = "084013",
    year = "2025"
}

@article{Volkel:2022aca,
    author = {V{\"o}lkel, Sebastian H. and Franchini, Nicola and Barausse, Enrico},
    title = "{Theory-agnostic reconstruction of potential and couplings from quasinormal modes}",
    eprint = "2202.08655",
    archivePrefix = "arXiv",
    primaryClass = "gr-qc",
    doi = "10.1103/PhysRevD.105.084046",
    journal = "Phys. Rev. D",
    volume = "105",
    number = "8",
    pages = "084046",
    year = "2022"
}

@article{Regge:1957td,
    author = "Regge, Tullio and Wheeler, John A.",
    title = "{Stability of a Schwarzschild singularity}",
    doi = "10.1103/PhysRev.108.1063",
    journal = "Phys. Rev.",
    volume = "108",
    pages = "1063--1069",
    year = "1957"
}

@article{Chandrasekhar:1975zza,
    author = "Chandrasekhar, S. and Detweiler, Steven L.",
    title = "{The quasi-normal modes of the Schwarzschild black hole}",
    doi = "10.1098/rspa.1975.0112",
    journal = "Proc. Roy. Soc. Lond. A",
    volume = "344",
    pages = "441--452",
    year = "1975"
}

@article{Vishveshwara:1970zz,
    author = "Vishveshwara, C. V.",
    title = "{Scattering of Gravitational Radiation by a Schwarzschild Black-hole}",
    doi = "10.1038/227936a0",
    journal = "Nature",
    volume = "227",
    pages = "936--938",
    year = "1970"
}

@article{Nollert:1999ji,
    author = "Nollert, Hans-Peter",
    title = "{TOPICAL REVIEW: Quasinormal modes: the characteristic `sound' of black holes and neutron stars}",
    doi = "10.1088/0264-9381/16/12/201",
    journal = "Class. Quant. Grav.",
    volume = "16",
    pages = "R159--R216",
    year = "1999"
}

@article{Kokkotas:1999bd,
    author = "Kokkotas, Kostas D. and Schmidt, Bernd G.",
    title = "{Quasinormal modes of stars and black holes}",
    eprint = "gr-qc/9909058",
    archivePrefix = "arXiv",
    doi = "10.12942/lrr-1999-2",
    journal = "Living Rev. Rel.",
    volume = "2",
    pages = "2",
    year = "1999"
}

@article{Berti:2009kk,
    author = "Berti, Emanuele and Cardoso, Vitor and Starinets, Andrei O.",
    title = "{Quasinormal modes of black holes and black branes}",
    eprint = "0905.2975",
    archivePrefix = "arXiv",
    primaryClass = "gr-qc",
    doi = "10.1088/0264-9381/26/16/163001",
    journal = "Class. Quant. Grav.",
    volume = "26",
    pages = "163001",
    year = "2009"
}

@article{Sanchez:1976xm,
    author = "Sanchez, Norma G.",
    title = "{The Wave Scattering Theory and the Absorption Problem for a Black Hole}",
    reportNumber = "PRINT-76-0510 (MEUDON)",
    doi = "10.1103/PhysRevD.16.937",
    journal = "Phys. Rev. D",
    volume = "16",
    pages = "937--945",
    year = "1977"
}

@article{Sanchez:1977si,
    author = "Sanchez, Norma G.",
    title = "{Absorption and Emission Spectra of a Schwarzschild Black Hole}",
    reportNumber = "Print-77-0242 (MEUDON)",
    doi = "10.1103/PhysRevD.18.1030",
    journal = "Phys. Rev. D",
    volume = "18",
    pages = "1030",
    year = "1978"
}

@article{Page:1976df,
    author = "Page, Don N.",
    title = "{Particle Emission Rates from a Black Hole: Massless Particles from an Uncharged, Nonrotating Hole}",
    doi = "10.1103/PhysRevD.13.198",
    journal = "Phys. Rev. D",
    volume = "13",
    pages = "198--206",
    year = "1976"
}

@article{Schutz:1985km,
    author = "Schutz, Bernard F. and Will, Clifford M.",
    title = "{BLACK HOLE NORMAL MODES: A SEMIANALYTIC APPROACH}",
    reportNumber = "PRINT-85-0063 (WASH.U.,ST.LOUIS)",
    doi = "10.1086/184453",
    journal = "Astrophys. J. Lett.",
    volume = "291",
    pages = "L33--L36",
    year = "1985"
}

@article{Iyer:1986np,
    author = "Iyer, Sai and Will, Clifford M.",
    title = "{Black Hole Normal Modes: A {WKB} Approach. 1. Foundations and Application of a Higher Order {WKB} Analysis of Potential Barrier Scattering}",
    reportNumber = "Print-86-1482 (WASH. U., ST. LOUIS)",
    doi = "10.1103/PhysRevD.35.3621",
    journal = "Phys. Rev. D",
    volume = "35",
    pages = "3621",
    year = "1987"
}

@article{Iyer:1986nq,
    author = "Iyer, Sai",
    title = "{BLACK HOLE NORMAL MODES: A WKB APPROACH. 2. SCHWARZSCHILD BLACK HOLES}",
    reportNumber = "Print-86-1483 (WASH. U., ST. LOUIS)",
    doi = "10.1103/PhysRevD.35.3632",
    journal = "Phys. Rev. D",
    volume = "35",
    pages = "3632",
    year = "1987"
}

@article{Leaver:1985ax,
    author = "Leaver, E. W.",
    title = "{An Analytic representation for the quasi normal modes of Kerr black holes}",
    doi = "10.1098/rspa.1985.0119",
    journal = "Proc. Roy. Soc. Lond. A",
    volume = "402",
    pages = "285--298",
    year = "1985"
}

@article{Konoplya:2023moy,
    author = "Konoplya, R. A. and Zhidenko, A.",
    title = "{Analytic expressions for quasinormal modes and grey-body factors in the eikonal limit and beyond}",
    eprint = "2309.02560",
    archivePrefix = "arXiv",
    primaryClass = "gr-qc",
    doi = "10.1088/1361-6382/ad0a52",
    journal = "Class. Quant. Grav.",
    volume = "40",
    number = "24",
    pages = "245005",
    year = "2023"
}

@article{Malik:2024cgb,
    author = "Malik, Zainab",
    title = "{Correspondence between quasinormal modes and grey-body factors for massive fields in Schwarzschild-de~Sitter spacetime}",
    eprint = "2412.19443",
    archivePrefix = "arXiv",
    primaryClass = "gr-qc",
    doi = "10.1088/1475-7516/2025/04/042",
    journal = "JCAP",
    volume = "04",
    pages = "042",
    year = "2025"
}

@article{Skvortsova:2024msa,
    author = "Skvortsova, Milena",
    title = "{Quantum corrected black holes: testing the correspondence between grey-body factors and quasinormal modes}",
    eprint = "2411.06007",
    archivePrefix = "arXiv",
    primaryClass = "gr-qc",
    doi = "10.1140/epjc/s10052-025-14589-w",
    journal = "Eur. Phys. J. C",
    volume = "85",
    number = "8",
    pages = "854",
    year = "2025"
}

@article{Bolokhov:2024otn,
    author = "Bolokhov, S. V. and Skvortsova, Milena",
    title = "{Correspondence between quasinormal modes and grey-body factors of spherically symmetric traversable wormholes}",
    eprint = "2412.11166",
    archivePrefix = "arXiv",
    primaryClass = "gr-qc",
    doi = "10.1088/1475-7516/2025/04/025",
    journal = "JCAP",
    volume = "04",
    pages = "025",
    year = "2025"
}

@article{Dubinsky:2024vbn,
    author = "Dubinsky, Alexey",
    title = "{Gray-body factors for gravitational and electromagnetic perturbations around Gibbons{\textendash}Maeda{\textendash}Garfinkle{\textendash}Horowitz{\textendash}Strominger black holes}",
    eprint = "2412.00625",
    archivePrefix = "arXiv",
    primaryClass = "gr-qc",
    doi = "10.1142/S0217732325501111",
    journal = "Mod. Phys. Lett. A",
    volume = "40",
    number = "28",
    pages = "2550111",
    year = "2025"
}

@article{Han:2025cal,
    author = "Han, Hyewon and Gwak, Bogeun",
    title = "{Correspondence between quasinormal modes and greybody factors in five-dimensional black holes}",
    eprint = "2508.12989",
    archivePrefix = "arXiv",
    primaryClass = "gr-qc",
    doi = "10.1103/n2ns-drkp",
    journal = "Phys. Rev. D",
    volume = "113",
    number = "6",
    pages = "064058",
    year = "2026"
}

@article{Han:2026fpn,
    author = "Han, Hyewon and Gwak, Bogeun",
    title = "{Correspondence between quasinormal modes and grey-body factors of Schwarzschild--Tangherlini black holes}",
    eprint = "2601.18613",
    archivePrefix = "arXiv",
    primaryClass = "gr-qc",
    month = "1",
    year = "2026"
}

@article{Huang:2025rxx,
    author = "Huang, Zun-Xian and Li, Peng-Cheng",
    title = "{Quasinormal mode/grey-body factor correspondence for Kerr black holes}",
    eprint = "2512.23510",
    archivePrefix = "arXiv",
    primaryClass = "gr-qc",
    month = "12",
    year = "2025"
}

\end{document}